\documentclass[aps,physrev,superscriptaddress,notitlepage,twocolumn]{revtex4-2}

\usepackage{graphicx}
\usepackage{amsmath,amssymb,amsfonts}
\usepackage{braket}
\usepackage{comment}
\usepackage{xcolor}
\usepackage{url}
\usepackage[hidelinks]{hyperref}
\usepackage{natbib}
\usepackage{dcolumn}
\usepackage{bm}

\usepackage{lineno}
\usepackage[normalem]{ulem}

\newcommand{\JQI}{Joint Quantum Institute and Joint Center for Quantum Information and Computer Science, University of Maryland and NIST, College Park, Maryland 20742, USA}

\newcommand{\DukeA}{Duke Quantum Center, Duke University, Durham, NC 27701}
\newcommand{\DukeB}{Department of Electrical and Computer Engineering, Duke University, Durham, NC 27708}
\newcommand{\DukeC}{Department of Physics, Duke University, Durham, NC 27708}
\newcommand{\IonQ}{IonQ, Inc., College Park, MD 20740}

\begin{document}

\title{Engineering dynamically decoupled quantum simulations with trapped ions}

\author{W. Morong}
\email{current address: AWS Center for Quantum Computing, Pasadena, California 91125, USA. Work done prior to joining Amazon.}
\affiliation{\JQI}

\author{K. S. Collins}
\email{ksc64@umd.edu}
\affiliation{\JQI}

\author{A. De}
\affiliation{\JQI}

\author{E. Stavropoulos}
\affiliation{\DukeA}
\affiliation{\DukeC}

\author{T. You}
\affiliation{\DukeA}
\affiliation{\DukeB}

\author{C. Monroe}
\affiliation{\JQI}
\affiliation{\DukeA}
\affiliation{\DukeB}
\affiliation{\DukeC}
\affiliation{\IonQ}

\begin{abstract}
    An external drive can improve the coherence of a quantum many-body system by averaging out noise sources. It can also be used to realize models that are inaccessible in the static limit, through Floquet Hamiltonian engineering. The full possibilities for combining these tools remain unexplored. We develop the requirements needed for a pulse sequence to decouple a quantum many-body system from an external field without altering the intended dynamics. Demonstrating this technique experimentally in an ion-trap platform, we show that it can provide a large improvement to coherence in real-world applications. Finally, we engineer an approximate quantum simulation of the Haldane-Shastry model, an exactly solvable paradigm for long-range interacting spins. Our results expand and unify the quantum simulation toolbox.
\end{abstract}

\maketitle

\clearpage

\section{Introduction}
\label{sec:intro}

The past few decades have seen a steady increase in the ability to control quantum states of simple systems, driven by improvements in maintaining the coherence of these systems. The most direct way to do this is to reduce their coupling to the environment through better engineering, which may involve materials developments or active stabilization of key parameters. However, since the development of NMR spin manipulation, a second strategy has also been known: improving the coherence of a system not in a fundamental way, but by strategically driving it with global operations that cancel out known and relatively slowly-varying noise sources \cite{Hahn1950, Viola1999, Souza2012}. This strategy, known as dynamical decoupling, has been extensively developed as a technique to reduce the coupling of quantum systems to external fields and thereby improve their internal coherences, which may allow them to serve as quantum memories \cite{Biercuk2009a,DeLange2010,Wang2021}. Similarly, modern quantum computers often apply dynamical decoupling sequences to qubits that remain idle for part of a computation \cite{Pokharel2018}. In both cases, one generally wants to completely cancel out any influence on the system.

More recently, the dynamics of interacting quantum systems under a periodic external drive have themselves become a subject of extensive investigation. A primary example of this is the study of discrete time crystals \cite{Else2019,Khemani2019,Zhang2017a,Choi2017a,Kyprianidis2021,Randall2021,Mi2022}. Other recent works have explored Floquet Hamiltonian engineering: the ability to use a periodic driving sequence to design a desired effective Hamiltonian that the system does not natively realize \cite{Hayes2014, Choi2020, Geier2021,Scholl2022, Kranzl2022}. 

Here, we combine these subjects to apply dynamical decoupling to a tunable quantum simulator. Although this technique is broadly applicable, we focus on its implementation with trapped ions, which are a leading quantum simulation platform \cite{Blatt2012, Monroe2019a}. This extends previous work in two ways. First, dynamical decoupling is applied not to preserve a quantum state, but to preserve evolution under a target Hamiltonian (or Floquet unitary operator). This adds the constraint, typically not present, that the decoupling pulse sequence not alter the desired dynamics. Demonstrating this on a trapped-ion simulator, we show that the coherence time can be substantially improved, in some cases by an order of magnitude. Second, while dynamical decoupling is often applied to a system with fixed intrinsic interactions, in a trapped-ion simulator there is significant freedom in the form of the Hamiltonian. The ability to periodically change the Hamiltonian in sync with global pulses opens up new possibilities for decoupling sequences, and for the combination of these with Hamiltonian engineering techniques. As a demonstration, we engineer an approximate decoupled quantum simulation of the Haldane-Shastry model, with dynamics reflecting key properties of this model including integrability \cite{Haldane1988,Shastry1988}.

This paper is organized as follows: Section~\ref{sec:pulses} describes the general requirements for a decoupling pulse sequence to eliminate certain noise terms without affecting overall dynamics. Section~\ref{sec:ions} gives a brief overview of trapped-ion quantum simulation, with a particular focus on the primary sources of decoherence. With these preliminaries in hand, Section~\ref{sec:DDexamples} develops example dynamical decoupling and Hamiltonian engineering sequences for this trapped ion system and Section~\ref{sec:twoIonTests} presents tests with two ions validating the effectiveness of these sequences. Section~\ref{sec:multiIonTests} extends this to larger systems, including an approximate realization of the Haldane-Shastry model incorporating dynamical decoupling and Floquet engineering. Finally, Section~\ref{sec:outlook} gives a summary and outlook.

\section{Principles of dynamical decoupling}

\label{sec:pulses}

\begin{figure*}[!htb]
\centering
\includegraphics[width= 0.8 \textwidth]{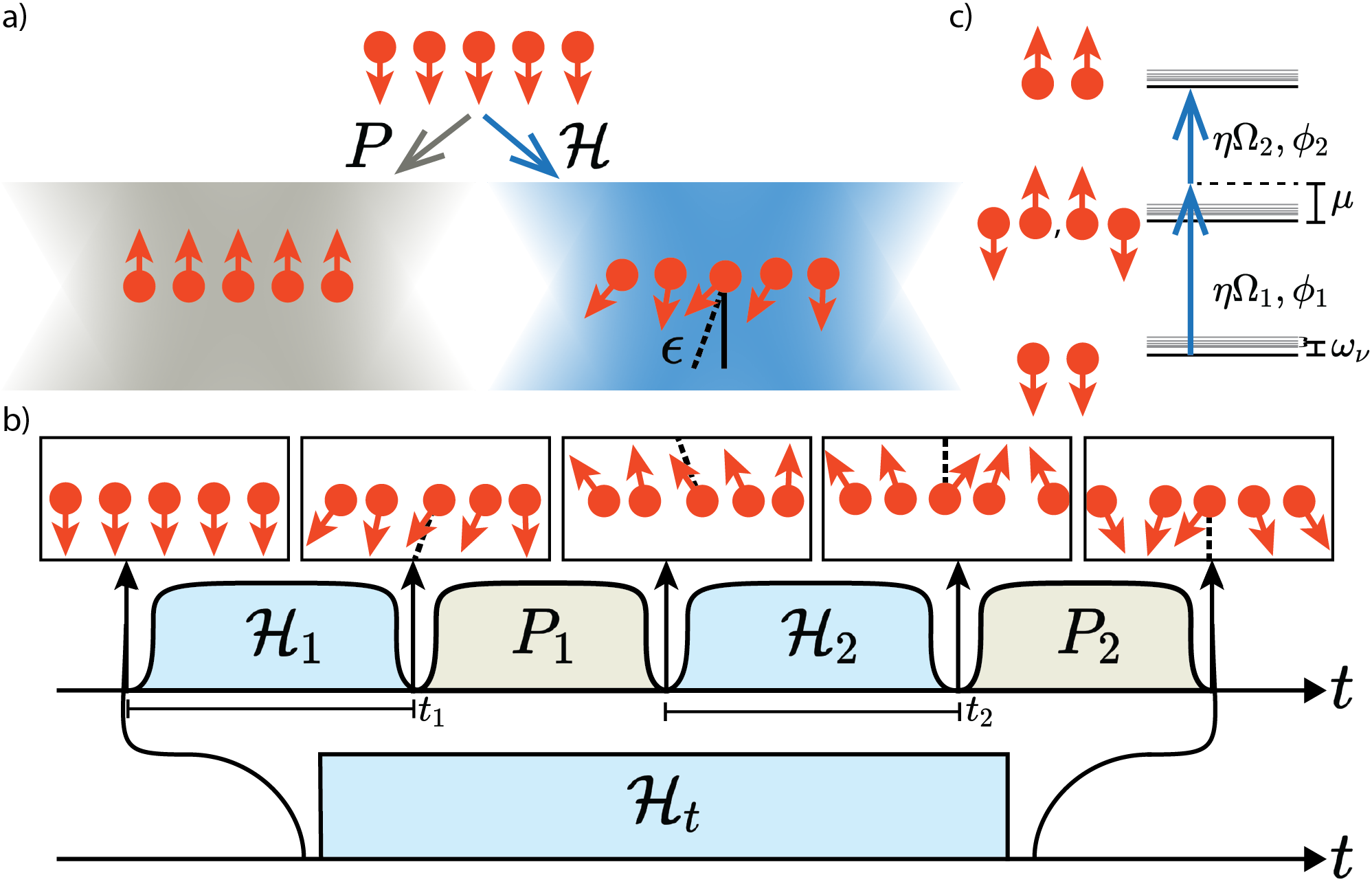}
\caption{Schematic of the dynamical decoupling scheme for quantum simulations. a) We consider a spin system that can be subjected to two classes of controllable operations (illustrated as driven by a pair of laser beams): overall global rotation pulses ($P_n$), and Hamiltonian evolution for some period of time ($\mathcal{H}_n$). The Hamiltonian evolution induces interactions (represented by fluctuating spin angles), but can also introduce a coherent error (represented by a tilt of $\epsilon$ in the average spin axis). b) The goal of the dynamical decoupling sequence is to alternate periods of $\mathcal{H}_n$ and $P_n$ operations, so that the final evolution after the pulse sequence corresponds to evolution (up to an overall rescaling) under a target Hamiltonian $\mathcal{H}_{t}$ while the error $\epsilon$ is cancelled out. c) Level schematic for spin-spin interactions generated by the M\o{}lmer-S\o{}rensen technique described in Sec.~\ref{sec:ions}. For simplicity, the two-ion case is shown and only transitions from the absolute ground state are drawn.}
\label{fig:Fig0}
\end{figure*}

We analyze dynamical decoupling sequences using the framework of Average Hamiltonian Theory, which concerns the average Hamiltonian $\overline{\mathcal{H}}$ that approximately describes slow evolution under some pulse sequence \cite{Brinkmann2016}. For a periodic (Floquet) sequence, this is equivalent to the static Hamiltonian resulting from a series expansion (the Floquet-Magnus expansion) in the drive frequency \cite{Kuwahara2016}. Our goal will be to design a pulse sequence that makes $\overline{\mathcal{H}}$ as similar as possible to the target Hamiltonian for the simulation, $\mathcal{H}_t$. Following previous expositions of Average Hamiltonian Theory, we consider the Hamiltonian in an interaction picture in which the operators act in a rotated frame determined by the previous pulses, which is often called the ``toggling frame''. For a pulse sequence consisting of evolution under Hamiltonians $\mathcal{H}_1$ to $\mathcal{H}_n$ for times $t_1$ to $t_n$, which are each followed by pulses $P_1$ to $P_n$ (see Fig.~\ref{fig:Fig0}), the unitary evolution operator $(\hbar =1)$ is

\begin{align}
U_1&=P_n e^{-i \mathcal{H}_n t_n}\cdots P_1 e^{-i \mathcal{H}_1 t_1}\\
&=(P_n\cdots P_1)e^{-i(\mathcal{H}'_n t_n)}\cdots e^{-i(\mathcal{H}'_1 t_1)},
\label{eq:PulseSequence}
\end{align}
resulting in the toggling-frame Hamiltonians $\mathcal{H}'_n:$
\begin{equation}
\mathcal{H}'_{n}=(P_{n-1}\cdots P_1)^{-1}\mathcal{H}_{n}(P_{n-1}\cdots P_1).
\end{equation}

A basic result of Average Hamiltonian Theory is that the resulting dynamics are approximately described by evolution according to the following static Hamiltonian:

\begin{align}
U_1&\approx(P_n\cdots P_1)e^{-i\overline{\mathcal{H}} T}, \\
\overline{\mathcal{H}}&=\frac{\sum_n \mathcal{H}'_nt_n}{T},
\end{align}
with $T$ the total evolution time including the pulses. In words, the average Hamiltonian is, to lowest order in $|\mathcal{H}_n t_n|$, nothing more than the weighted average of the different rotated Hamiltonians that make up the sequence. Within this framework, we define a dynamically decoupled sequence for some operator $\mathcal{O}$, often an undesired noise term, as a pulse sequence which results in an average Hamiltonian that does not include $\mathcal{O}$, despite $\mathcal{O}$ appearing in at least one of the individual $\mathcal{H}_n$ of the sequence.

To design dynamical decoupling sequences that are suitable for a quantum simulation experiment, we usually impose two additional requirements on the pulse sequences. Most importantly, we require that, in the absence of noise (for $\mathcal{O}=0$), the unitary evolution is completely unchanged from that of $\mathcal{H}_t$, up to a known overall rescaling. To satisfy this, we also demand that the pulse sequence rotate the frame back to the original frame at the end of a cycle. These conditions are:

\begin{align}
     \mathcal{H}'_n &=\mathcal{H}_{t}, \\
    P_n \cdots P_1&=I.
\end{align}

As a result, in the absence of noise, the Average Hamiltonian Theory description of the dynamics is \emph{exact}. Higher-order corrections to $\overline{\mathcal{H}}$ are determined by the commutators $[\mathcal{H}'_n,\mathcal{H}'_m]$, which all vanish. Consequently, provided that the pulse sequence is fast enough to effectively implement decoupling, the dynamics are guaranteed to follow the target Hamiltonian. While this is highly advantageous for many applications, in Sec.~\ref{sec:Heisenberg} we explore the possibility of removing this constraint to engineer average Hamiltonians that are not available in the static limit.

As a lesser constraint, we limit ourselves to decoupling sequences that are periodic in time. Accordingly, $U_1$ becomes the unitary evolution under one cycle, with the total evolution after $m$ steps being $U=\Pi_m (U_1)^m$. Combined with the previous condition, this allows us to vary the total evolution time by integer cycles and observe time evolution matching the target Hamiltonian, as long as decoupling is effective.

While simple dynamical decoupling sequences can often be designed by inspection, more elaborate sequences can be systematized in several ways. Previous work \cite{Choi2020} has developed an approach to determining $\overline{\mathcal{H}}$ based on a pulse-sequence matrix representation, focusing on the case of a fixed $\mathcal{H}_n$. The Floquet-Magnus expansion can also be used to calculate higher-order terms in $\overline{\mathcal{H}}$ \cite{Kuwahara2016} (see Appendix~\ref{sec:AppendixError}), which can be suppressed at the cost of longer, symmetrized pulse sequences. Finally, we note that allowing non-periodic pulse sequences can be advantageous for cancelling certain types of noise \cite{Biercuk2009a}, although such sequences may complicate the interpretation of dynamics with a varying evolution time.

We now provide a brief description of how the $\mathcal{H}_n$ and $P_n$ operations are generated in our trapped-ion platform, and apply this framework to the specifics of this system.

\section{Quantum simulation with trapped ions}
\label{sec:ions}

Here we give an overview of the generation of long-range Ising-like Hamiltonians using the M\o{}lmer-S\o{}rensen scheme \cite{Molmer1999}, following previous treatments \cite{James2007, Blatt2012, Monroe2019a}. We consider a linear chain of $N$ ions in a radiofrequency (rf) Paul trap \cite{Leibfried2003}, which is uniformly illuminated by multiple optical tones nearly resonant with a series of joint spin and motional transitions (often referred to as red and blue sideband transitions). Tone $\lambda$ generates a Hamiltonian term of the following form:

\begin{align}
    \mathcal{H}_\lambda(t) &= \sum_j\sigma^+_j\left[\frac{\Omega_\lambda}{2}(1+i\sum_\nu \eta b_{\nu,j} (a_\nu e^{-i\omega_\nu t} + a^\dagger_\nu e^{i\omega_\nu t}))\right.\nonumber \\
    &\times e^{-i\mu_\lambda t+i\phi_\lambda} \biggr]+H.c. 
    \label{eq:Molmer}
\end{align}
Here $j$ is the ion index and $\nu$ is the normal mode index, $a_\nu$ is the destruction operator of a phonon of motion for a normal mode of the ion chain, $\sigma_j^+$ is a raising operator for the two-state pseudospin (or qubit), $\eta$ is the Lamb-Dicke parameter, $b_{\nu,j}$ is the mode $\nu$ amplitude for ion $j$, $\omega_\nu$ is the mode frequency, and $\Omega_\lambda$, $\mu_\lambda$, and $\phi_\lambda$ are the Rabi rate, detuning from the qubit transition, and phase of tone $\lambda$, respectively. We have taken the Lamb-Dicke limit, implying that $\eta \ll 1$. These tones can be used to generate a family of Hamiltonians, of which we briefly describe the most relevant cases.

First, with two tones that are symmetrically detuned near the motional transitions, $\mu_1=-\mu_2=\mu$, the evolution can be analyzed using the Magnus expansion and shown over long times to approximately follow Hamiltonian evolution with the following form \cite{Monroe2019a}:

\begin{align}
    \label{eq:FullMSHamiltonian}
    \mathcal{H}&=\sum_{j<j',\nu}\frac{i\omega_\nu \eta^2\Omega_1\Omega_2 b_{\nu,j} b_{\nu,j'}}{\mu^2-\omega_\nu^2}(\cos\phi_s\sigma^x_j-\sin\phi_s\sigma^y_j)\nonumber \\ &\times(\cos\phi_s\sigma^x_{j'}-\sin\phi_s\sigma^y_{j'})+\sum_j \frac{(\Omega_2^2-\Omega_1^2)}{4\mu}\sigma^z_j ,
\end{align}
where $\phi_s=(\phi_1+\phi_2+\pi)/2$ (see Fig.~\ref{fig:Fig0}c). The first term, describing an effective spin-spin interaction, can be approximately written (for $\mu>0$) as an Ising-like term with a power-law spatial distribution that is tuned by $\mu$, while the second term contains the AC Stark shifts (or light shifts) of the qubit levels from the two frequencies that vanish when they have balanced strengths.

Second, with one tone of $\mu=0$, this Hamiltonian (Eq.~\ref{eq:Molmer}) may be simplified to an effective magnetic field term:

\begin{align}
    \mathcal{H} &= \sum_j\left[\frac{\Omega}{2} (\cos{\phi}\sigma^x_j-\sin{\phi}\sigma^y_j) \right].
\end{align}
Such a tone may be applied in isolation, to realize global spin rotations, or in combination with an interaction term.

In addition to these directly generated Hamiltonian terms, an effective $B^z$ field can be implemented by means of a rotating frame transformation that redefines the qubit frequency. In Appendix~\ref{sec:AppendixBz} we describe how to combine this continuous rotating-frame transformation with the toggling-frame from the pulses to use this technique in dynamical decoupling sequences.

A common target Hamiltonian for quantum simulations is a long-range transverse Ising model, such as:

\begin{align}
\mathcal{H}=\sum_{j<j'}J_{j,j'}\sigma^x_j\sigma^x_{j'}+\sum_j B^y\sigma_j^y .
\end{align}
However, several errors can cause the observed dynamics to deviate from evolution according to this Hamiltonian. These errors include:
\begin{itemize}
    \item Fluctuations in the ratio of $\Omega_1/\Omega_2$ generating the interactions, causing the AC Stark term to drift away from zero. This is of special concern because the AC Stark shifts, unlike the spin-spin couplings, are not suppressed by $\eta^2$ (see Eq.~\ref{eq:FullMSHamiltonian}). As a result, they are often much larger than the intended terms in the Hamiltonian (typically about an order of magnitude larger than $J_{j,j+1}$ for our experimental parameters) and therefore require a high degree of fractional stability.
    \item More generally, fluctuations in any of the laser intensities generating the Hamiltonian terms, causing the values of $J_{j,j'}$ or $B^y$ to vary over time.
    \item Errors arising from finite population of the motional states. This can occur from direct excitation of these transitions by the tones, or due to heating processes that these tones couple into the spin dynamics (see Appendix~\ref{sec:AppendixNoise}) \cite{Monroe2019a}.
\end{itemize}
In the following sections we demonstrate dynamical decoupling sequences that suppress the AC Stark noise, and show that this leads to a large increase in coherence for sufficiently large detuning $\mu$.

\section{Application to trapped ions and example sequences}
\label{sec:DDexamples}
We now apply dynamical decoupling to an experimental trapped ion simulator, in which noise enters as a (possibly site-dependent) effective $B^z$ field: $\mathcal{O}=\sum_j\epsilon_j(t)\sigma^z_j$. Several features particular to this experimental platform prove to be especially convenient for this goal. First, the noise varies relatively slowly compared to the duration of pulses. Second, both the pulse operations and Hamiltonian terms are generated by laser tones, and can be turned on and off or modified individually or in concert. Finally, the noise is only present when the interactions are on, and is approximately independent of the details of the Hamiltonian.

Two consequences of this control are worth highlighting. First, there is essentially no unwanted evolution during the pulses. This can be contrasted with many natural systems with fixed interactions, in which there is some undesired evolution from the interactions during the finite lengths of the decoupling pulses. Second, while the average Hamiltonian that results from a pulse sequence applied to a fixed Hamiltonian is constrained by the symmetry group associated with the pulses \cite{Viola1999}, the ability in this system to turn on and off different Hamiltonians in sync with the pulses allows us to circumvent this limitation and realize average Hamiltonians with arbitrary symmetries. This is demonstrated explicitly in the following section.

It is useful to explicitly provide the transformations of the Pauli operators under global $\pi$ pulses. These may be summarized as:

\begin{equation}
R_k^{-1}(\pi) \sigma^{k'}_{j'} R_k(\pi)=(-1)^{1+\delta_{k,k'}}\sigma^{k'}_{j'}.
\end{equation}
where $k,k'=\{x,y,z\}$ index the Pauli operators (excluding the identity) and $R_k(\theta)=e^{-i\sum_j\theta\sigma^k_j/2}$. Geometrically, a $\pi$ rotation about axis $k$ flips the sign of a Pauli operator unless the axis of rotation and the Pauli operator direction coincide. This allows for a natural categorization of operators based on their parity under a given $\pi$ rotation. For example, under the rotation $R_y(\pi)$, $\sigma^x_j$ is odd but $\sigma^x_j \sigma^x_{j'}$ is even. Our strategy, similar to earlier proposals for high-fidelity quantum gates \cite{Duan1999,Piltz2013,Dong2020,Zhang2022, Valahu2022}, is to use this difference to approximately invert the part of the unitary operator corresponding to evolution from the noise, undoing this evolution over two interaction periods separated by a $\pi$ pulse, while the desired evolution is unchanged. Looking back at the fundamental Hamiltonian (Eq.~\ref{eq:Molmer}), this is possible because although it generically has single spin operators with odd parities, the overall time dependence means that changing the sign is not equivalent to inverting the unitary. Instead, the M\o{}lmer-S\o{}rensen interaction relies on a geometric phase accumulated over the ions' spin-motion trajectory \cite{Kim2011}, which is invariant under a $\pi$ rotation.

\subsection{Example 1: CPMG sequence}
\label{sec:CPMG}
As a minimal example of a decoupling sequence satisfying our requirements, we consider a typical asymmetric Carr-Purcell-Meiboom-Gill (CPMG) decoupling sequence \cite{Souza2012}. The target Hamiltonian includes spin-spin interactions and uniform fields along each direction:

\begin{equation}
    \mathcal{H}_{t}=\sum_{j<j'}J_{j,j'}\sigma_j^x \sigma_{j'}^x+\sum_j B^x\sigma^x_j+B^y\sigma^y_j+B^z\sigma^z_j.
    \label{eq:targetHamCPMG}
\end{equation}
For this sequence $n=2$ (of Eq.~\ref{eq:PulseSequence}), $t_1=t_2$, and both pulses are $\pi$ pulses along one axis taken to be $y$. A single unit of the evolution is:

\begin{align}
     U_1 &= R_y(\pi)  e^{-i \mathcal{H}_2 t_1}   R_y(\pi)   e^{-i \mathcal{H}_1 t_1} , \\
     \mathcal{H}_1&=\sum_{j<j'}J_{j,j'}\sigma_j^x \sigma_{j'}^x+\sum_j\left( B^x\sigma^x_j+B^y\sigma^y_j\right.\nonumber\\
     &\left.+B^z\sigma^z_j+\epsilon_j(t)\sigma^z_j\right),  \\
     \mathcal{H}_2&=\sum_{j<j'}J_{j,j'}\sigma_j^x \sigma_{j'}^x+\sum_j \left((-B^x)\sigma^x_j+ B^y\sigma^y_j\right.\nonumber\\
     &\left.+(-B^z)\sigma^z_j+\epsilon_j(t)\sigma^z_j\right).
\end{align}
Note that we have explicitly included the noise term $\mathcal{O}=\epsilon_j(t)\sigma^z_j$ in both $\mathcal{H}_1$ and $\mathcal{H}_2$.

This sequence satisfies our requirements, with the terms in the target Hamiltonian remaining unchanged for different reasons. The spin-spin interaction term has even parity under any $\pi$ pulse, and is therefore unchanged by $R_y(\pi)$, the $B^y$ term has even parity under $R_y(\pi)$ because their axes are aligned, and finally the $B^x$ and $B^z$ terms have odd parity under $R_y(\pi)$, but are correspondingly flipped between $\mathcal{H}_1$ and $\mathcal{H}_2$ to undo their transformation. As a result, the evolution under the target Hamiltonian is unchanged by the pulses, only occurring in a rotated frame that returns to the initial frame after both $\pi$ pulses. However, the $\sigma^z_j$ noise, which has odd parity and does not rotate with the pulse frame, reverses sign after each $\pi$ pulse and is approximately cancelled. The exact degree of suppression depends on the power spectrum of $\epsilon(t)$ compared to the pulse frequency \cite{Biercuk2009a}, as well as the commutator of the noise with the target Hamiltonian (see Appendix~\ref{sec:AppendixError}).

As promised in the introduction to this section, this example demonstrates that the ability to vary the Hamiltonian during the pulse sequence has expanded the range of possibilities for $\overline{\mathcal{H}}$. If we were limited to a single Hamiltonian, for example $\mathcal{H}_1$, it would not be possible to have terms proportional to $\sigma^x_j$ or $\sigma^z_j$ in $\overline{\mathcal{H}}$, because they do not respect the symmetry of the pulse sequence. Thus, controllability of the Hamiltonian has significantly expanded the range of models amenable to dynamical decoupling.

\subsection{Example 2: XY sequence}
The second sequence we consider is a version of the XY dynamical decoupling sequence \cite{Souza2012} for the same $\mathcal{H}_{t}$ (Eq.~\ref{eq:targetHamCPMG}):

\begin{align}
     U_1 &= R_y(\pi) e^{-i \mathcal{H}_4 t_1} R_x(\pi)  e^{-i \mathcal{H}_3 t_1} \nonumber\\
     &\times R_y(\pi) e^{-i \mathcal{H}_2 t_1}  R_x(\pi)  e^{-i \mathcal{H}_1 t_1}. 
\end{align}
This sequence traverses a more complex rotating frame that only returns to itself after four global rotations. As a result, the signs of the $B$ field terms must be cycled in the following way from $\mathcal{H}_1$ to $\mathcal{H}_4$: $(+B_x,+B_y,+B_z)\rightarrow(+B_x,-B_y,-B_z)\rightarrow(-B_x,-B_y,+B_z)\rightarrow(-B_x,+B_y,-B_z)$, while the spin-spin term is unchanged for each Hamiltonian. While this has additional complexity relative to the CPMG sequence, it has the advantage of being insensitive to pulse-length errors \cite{Souza2012}, due to a chirality condition satisfied by the pulse sequence \cite{Choi2020}, and also decouples noise along the $x$ and $y$ directions in addition to $z$. A similar robustness to pulse-length errors can be achieved in the CPMG sequence by alternating the sign of the rotations between $R_y(\pi)$ and $R_y(-\pi)$.

\subsection{Example 3: Hamiltonian engineering a decoupled Heisenberg model}
\label{sec:Heisenberg}
Until now, we have imposed the stringent requirement that the Hamiltonian be exactly the same in each rotated frame. This ensures that the dynamics are unchanged by the pulse sequence, removing any higher-order terms in the Floquet-Magnus expansion. However, one can also relax this condition to engineer a desired average Hamiltonian that is not accessible in the static limit. For example, in a toggling frame set by $R_y(\pi/2)$ pulses, $\sigma^x_j \sigma^x_{j'}$ is rotated to $-\sigma^z_j \sigma^z_{j'}$, which is not otherwise accessible. Recently, this concept of Hamiltonian engineering has been used to extend the capability of quantum simulators \cite{Choi2020, Geier2021, Scholl2022,Kranzl2022}. While this imposes the extra condition that the cycle time (which can now be considered as a Trotter step \cite{Hatano2005,Childs2019}) be sufficiently small, within this limit it is natural to combine this technique with dynamical decoupling. For example, here is a pulse sequence that realizes, in the limit of a high-frequency drive, a dynamically decoupled long-range Heisenberg model ($\mathcal{H}_{t}=\sum_{j<j'} \ J_{j,j'}\vec{\sigma}_j\cdot\vec{\sigma}_{j'}/3)$ that is insensitive to pulse-length errors:

\begin{align}
    U_1 &=  R_y(-\pi/2)  e^{-i \mathcal{H}_{XX} t_1/2} R_y(\pi)  e^{-i \mathcal{H}_{XX} t_1/2} \nonumber\\
    &\times  R_y(\pi/2) e^{-i \mathcal{H}_{YY} t_1}  R_y(-\pi)  e^{-i \mathcal{H}_{XX} t_1}, 
    \label{eq:Heisenberg}
\end{align}
where $\mathcal{H}_{XX}=\sum_{j<j'}J_{j,j'}\sigma_j^x \sigma_{j'}^x$ and $\mathcal{H}_{YY}=\sum_{j<j'}J_{j,j'}\sigma_j^y \sigma_{j'}^y$. In Section~\ref{sec:Haldane-Shastry} we apply this sequence to experimentally engineer the Haldane-Shastry model.

\section{Two-ion tests of pulse sequence parameters}
\label{sec:twoIonTests}

\begin{figure}[!htb]
\centering
\includegraphics[width= 0.45 \textwidth]{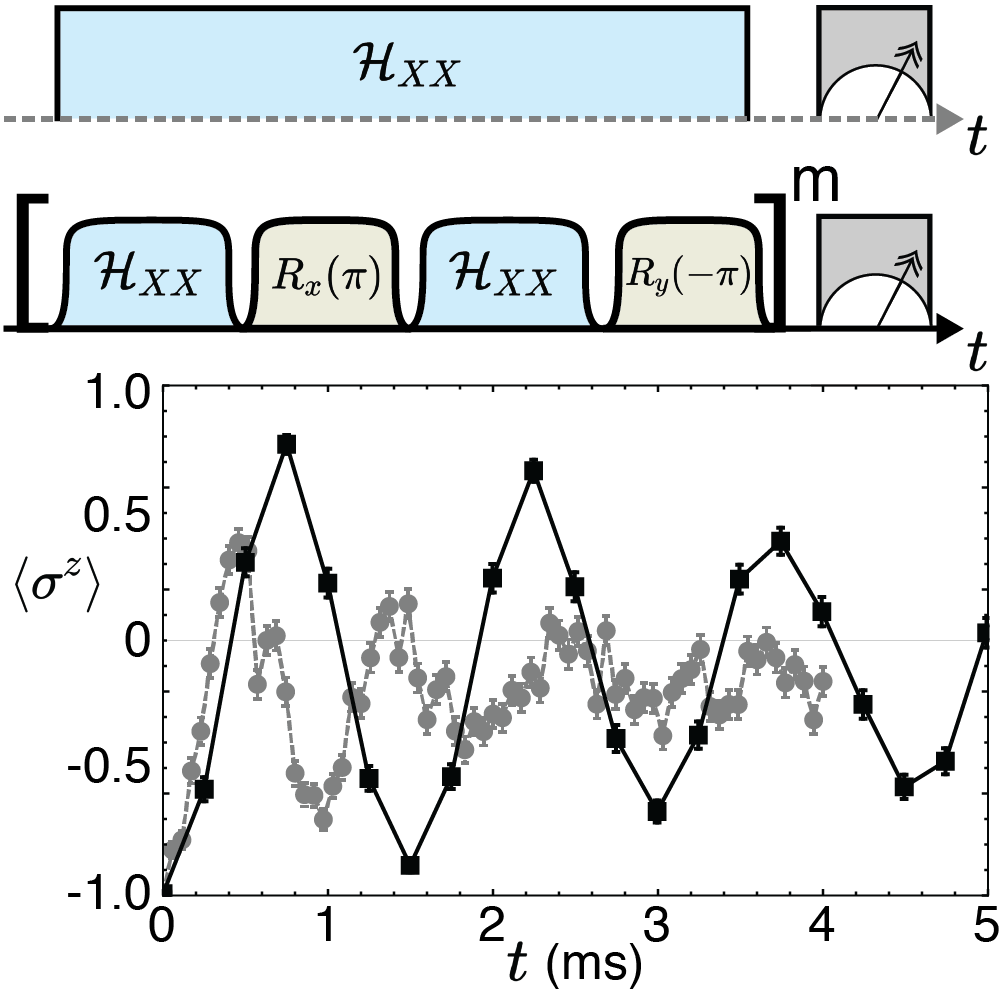}
\caption{Example of dynamical decoupling for a two-ion interaction. Direct evolution of the average ion magnetization, $\langle \sigma^z \rangle$, under $\mathcal{H}_{XX}$ (gray dashed line) is quickly damped due to slow dephasing noise from fluctuating Stark shifts. In comparison, the dynamical decoupling scheme (Eq.~\ref{eq:XYexpt}) consisting of alternating $\pi$ pulses about the $x$ and $y$ axes (black solid line) preserves coherence to substantially longer times, even after accounting for the slowing of the averaged interactions, without otherwise altering the intended dynamics. Each point represents the mean of 300 experimental repetitions, with error bars of 1 s.e.m. The experimental parameters used here are $t_p=20$ $\mu$s, $t_1=120$ $\mu$s, and $\delta/ \eta \Omega=5.0.$}
\label{fig:Fig1}
\end{figure}

In the following sections, we experimentally benchmark these pulse sequences. The experimental apparatus (see Appendix~\ref{sec:AppendixExperiment}) consists of 2--10 ${}^{171}$Yb$^+$ ions trapped in an rf Paul trap. We use two magnetic field-insensitive ground hyperfine states as the pseudospin $|\!\! \uparrow\rangle_z$ and $|\!\! \downarrow\rangle_z$ and perform coherent operations consisting of individual spin rotations, global spin rotations, and global long-range spin-spin interactions.

We initially optimize and benchmark the effectiveness of dynamical decoupling with minimal tests using two ions. The sequence is a simplified version of the XY sequence \footnote{We favor rotations about $-y$ rather than $y$ in experiment, as it corresponds to a phase of 90 degrees in our calculated waveform}:

\begin{equation}
     U_1 =  R_{-y}(\pi) e^{-i \mathcal{H}_{XX} t_1}  R_x(\pi) e^{-i \mathcal{H}_{XX} t_1}.
     \label{eq:XYexpt}
\end{equation}
Starting with both ions in $|\!\! \downarrow \downarrow\rangle_z$, the ideal state after application of $\mathcal{H}_{XX}(t)$ has a simple form: $|\psi(t)\rangle=\cos{(\overline{J_0}t)}|\!\! \downarrow \downarrow\rangle_z-i\sin{(\overline{J_0}t)}|\!\! \uparrow \uparrow\rangle_z$, where $\overline{J_0}$ is the spin-spin coupling of the averaged Hamiltonian $\overline{\mathcal{H}}$. Compared to the general form of the XY sequence in the previous section, both $\mathcal{H}_{XX}$ and the average ion $z$ magnetization $\langle \sigma^z \rangle=\sum_j \langle \sigma^z_j \rangle/N$ used here are symmetric under sign changes in $x$ and $y$, allowing us to use a fundamental time step with only two $\pi$ pulses, rather than four, without changing any measurement outcomes. Besides representing a basic building block of our quantum simulations, this sequence is also highly sensitive to the dephasing noise that we are trying to decouple.

Fig.~\ref{fig:Fig1} shows a typical example of this technique. The interspersed $\pi$ pulses effectively cancel out the slow AC Stark shift noise that is a primary cause of decoherence, while evolution under the target Hamiltonian is unaffected up to an overall scaling factor. Using data of this type, we quantify the success of the decoupling with the scaled coherence time $\overline{J_0}\tau$, the product of the coherence time $\tau$ with $\overline{J_0}$. More specifically, we fit the average magnetization $\langle \sigma^z\rangle$ to the function $g(t)=A(1-e^{t/\tau}\cos{(2\overline{J_0} t)})-1$. AC Stark shift noise leads both to finite $\tau$, due to dephasing, and to $A<1$, due to a nonzero average magnitude of the energy difference between the states $|\!\! \downarrow \downarrow\rangle_z$ and $|\!\! \uparrow \uparrow\rangle_z$.

For the data shown in Fig.~\ref{fig:Fig1}, the fit values (fits not shown) for $\overline{J_0}/(2\pi)$ are 0.52 kHz for bare evolution and 0.33 kHz with dynamical decoupling. The fit values for $\tau$ are 1.2 ms and 7.5 ms for bare evolution and with dynamical decoupling, respectively. This illustrates the central trade-off of these sequences: overall slower dynamics, but the potential for an increase in coherence time such that the product of the two improves (from $\overline{J_0}\tau=3.9$ to 15.6).

We now describe a few parameters of the dynamical decoupling sequence which must be chosen for optimal performance: the pulse shaping properties, the drive rate, and the detuning $\mu$ of the tones used to generate $\mathcal{H}_{XX}$.

\subsection{Pulse shaping}
We smoothly turn on and off both the interactions and the global rotations, to reduce unwanted motional transitions resulting from spectral broadening of the pulses. This is especially crucial for the interactions, which are relatively near-detuned from motional transitions. We use a Tukey window \cite{Zhang2017a} with shaping parameter $\alpha=2t_p/T$, where $t_p$, the pulse shaping time, describes the amount of time that the pulse is either being ramped up or down compared to total pulse length $T$. For the interactions, the most relevant quantity is $t_p \delta$, the product of the pulse shaping time, which determines the degree of broadening, and the detuning from the nearest motional resonance $\delta=(\mu-\omega_1)/2\pi$. We experimentally find that setting $t_p\delta \geq 3$ makes the induced error from motional excitations negligible, leading to a typical value of $t_p=20$ $\mu$s for our parameters. Global rotations, while less sensitive to pulse shape, are similarly given an $\alpha$ of 0.4.

The pulse shaping of the interactions leads to a decrease in the overall scale of the averaged Hamiltonian, which depends on the precise way that the rf waveform is imprinted onto the laser intensity. This can be theoretically predicted or determined with measurements of the ex-situ beam power, but we choose to characterize it with the in-situ ion dynamics. We find that the averaged spin-spin coupling for a single pulse of the interactions obeys $\beta=\int_0^{t_1} J (t) dt/(J_0 t_1)= (1-1.178t_p/t_1)$ (for $t_1>2t_p$), where $J(t)$ ramps from 0 to $J_0$. For our normal drive parameters, $\beta$ is near 0.8, meaning that the pulse shaping removes 20\% of the area under $J(t)$. The pulse-averaged $\beta J_0$ can then be used the standard formulas of Average Hamiltonian Theory. For example, in the XY sequence of Fig.~\ref{fig:Fig1}, the average Hamiltonian for two ions is:

\begin{equation}
    \overline{\mathcal{H}}=\overline{J_0} \sigma_1^x \sigma_{2}^x\text{,  }\overline{J_0}=(\beta J_0)\frac{t_1}{t_1+t_{\pi}},
    \label{eq:AvgJ0}
\end{equation}
simply reflecting the fraction of time during $U_1$ in which the interactions are applied.

\subsection{Drive rate}
\begin{figure}[!htb]
\centering
\includegraphics[width= 0.35 \textwidth]{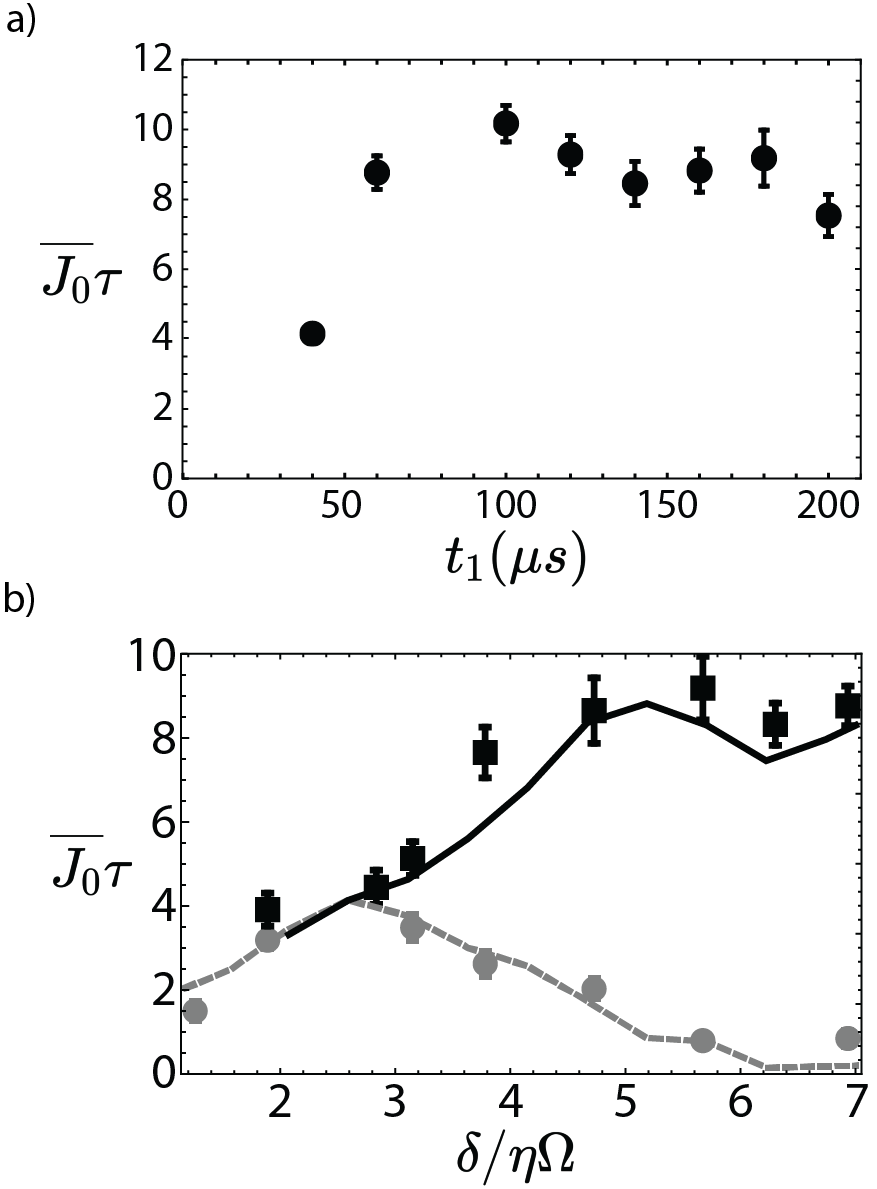}
\caption{Dependence of dynamical decoupling on pulse parameters. a): Dependence of dynamical decoupling on pulse length $t_1$. The scaled coherence time $\overline{J_0}\tau$ is largely flat for pulses longer than 50 $\mu$s, suggesting that the noise is slow enough to be largely cancelled out for any choice within this range. For this data, the detuning was fixed at $\delta/ \eta \Omega = 4.1$. For sufficiently fast drives, we see a decrease in $\overline{J_0}\tau$, which is caused by a decreasing $\overline{J_0}$ without any compensating improvement in $\tau$. b): Dependence of $\overline{J_0}\tau$ on detuning $\delta$, for both decoupled (black) and regular (gray) sequences. Points are experimental data, while solid lines are numerics incorporating the dominant noise sources (see Appendix~\ref{sec:AppendixNoise}). Decoupling is most effective in the regime of large detuning, $\delta/\eta\Omega > 5$, in which the evolution without decoupling is nearly completely destroyed. For this data, the drive was fixed at $t_1=120$ $\mu$s, and pulse shaping was fixed at $t_p=20$ $\mu$s throughout. Experimental data in both plots is averaged over 300 shots with error bars from the fit uncertainty.}
\label{fig:Fig3}
\end{figure}

The drive rate, or the value of $1/t_1$, is set by two considerations. The drive rate determines the bandwidth of noise that is suppressed, and must be chosen to be fast enough for the observed noise. The drive also sets the minimum time step, so it should be chosen so that the dynamics of interest are resolved. On the other hand, increasing the drive rate does not lead indefinitely to better performance because it slows down the experimental timescale, increasing sensitivity to any slow drifts that are not decoupled.

Varying our experimental drive rate over the range of 40-200 $\mu$s (Fig.~\ref{fig:Fig3}), we observe nearly no decrease in $\overline{J_0}\tau$ with increasing $t_1$, indicating that the noise we are cancelling out is mostly slower than the kHz scale. At sufficiently fast rates of driving, we see a decrease in $\overline{J_0}\tau$ because the fast drive leads to a reduction in the average Hamiltonian parameter $\overline{J_0}$ (see Eq.~\ref{eq:AvgJ0}) without a corresponding improvement in $\tau$.

\subsection{Detuning}

The M\o{}lmer-S\o{}rensen detuning $\delta$ is a key parameter in ion trap quantum simulations that determines the power-law range of the spin-spin coupling. It also controls the relative strengths of various sources of error. When $\delta$ is relatively small, coupling to the intermediate motional states is strong, and therefore errors associated with this motional coupling, such as fluctuations in the motional resonance frequencies, determine the coherence time. When $\delta$ is large, Stark shift noise becomes dominant, with the two being roughly equal near our normal point of operation of $\delta/\eta\Omega\approx 3$.

Fig.~\ref{fig:Fig3} shows the dependence of $\overline{J_0}\tau$ on $\delta$, for sequences with and without dynamical decoupling. At small $\delta$ the two are similar, while at large $\delta$ the dynamical decoupling begins to perform better. At the optimum value of $\delta/\eta\Omega \approx 5$, dynamical decoupling improves $\overline{J_0}\tau$ by over an order of magnitude. The dependence is captured well by a numerical model incorporating the relevant error sources (see Appendix~\ref{sec:AppendixNoise} and Fig.~\ref{fig:FigNoiseModel}). This suggests that this technique is especially valuable for quantum simulations requiring shorter-range interactions, which are reached with large $\delta$.

\section{Multi-ion tests}
\label{sec:multiIonTests}
\subsection{Dephasing test}

\begin{figure*}[!htb]
\centering
\includegraphics[width= 0.9 \textwidth]{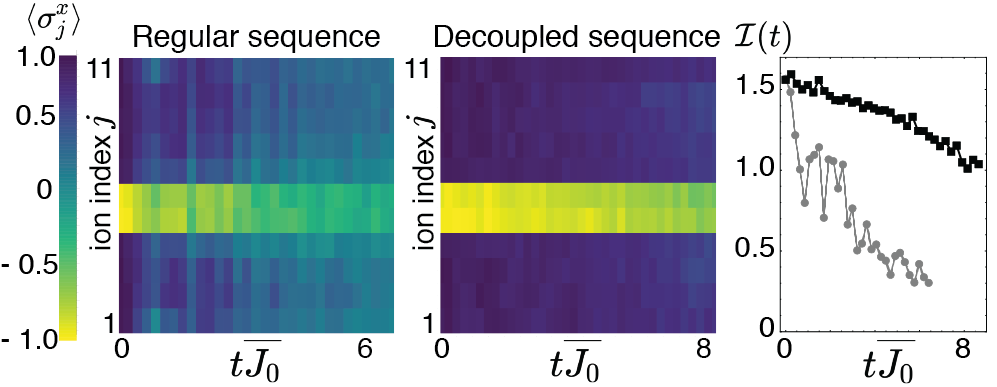}
\caption{Performance of dynamical decoupling to preserve trivial evolution under a many-body Hamiltonian. Ions are initialized as either up or down along $x$, and a long-range spin-spin Hamiltonian $\mathcal{H}_{XX}$ is applied for variable time, with and without CPMG decoupling pulses, before the $x$ magnetization is read out. In the absence of decoupling, stray AC Stark shifts induce dynamics and dephasing, while with decoupling these are suppressed. The parameters are $\delta/\eta\Omega=7.5$, $t_p=20$ $\mu$s, and $t_1=120$ $\mu$s. Evolution time is scaled by the nearest-neighbor spin-spin coupling averaged over the pulse sequence, $\overline{J_0}.$ Left panels: average magnetization of individual spins. Right panel: generalized imbalance $\mathcal{I}(t)$, measuring preservation of initial state (see text), for regular sequence (gray) and decoupled sequence (black). Points are the average of 500 experimental repetitions, with error bars smaller than the symbol size.}
\label{fig:Fig4}
\end{figure*}

It is desirable to check that these techniques extend to larger ion chains. This requires a change in the methodology for determining $\overline{J_0}$ and $\tau$, which were previously found by a fit to an analytical form for the two-ion dynamics. For the following multi-ion studies, we calibrate $\overline{J_0}$ using the M\o{}lmer-S\o{}rensen formula (Eq.~\ref{eq:FullMSHamiltonian}) and Average Hamiltonian Theory, while $\tau$ is estimated from an exponential fit to an observable of the system that is conserved by the ideal Hamiltonian but modified by decoherence. As a minimal demonstration, we have prepared a chain of eleven ions, whose spin-spin interactions obey an approximate long-range power law with a similar nearest-neighbor spin-spin coupling as in the two-ion tests (see Appendix~\ref{sec:AppendixExperiment}). The spins are individually initialized (with a single-spin addressing beam \cite{Lee2016a}) in a product state in which each spin has a definite value of $\sigma^x_j$, and evolved under a CPMG-type sequence:

\begin{equation}
     U_1 =  R^{\text{BB1}}_{-y}(\pi) e^{-i \mathcal{H}_{XX} t_1}  R^{\text{BB1}}_{-y}(\pi)  e^{-i \mathcal{H}_{XX} t_1} .
\end{equation}
To avoid errors associated with the rotations, we replace each bare rotation $R_{-y}(\pi)$ with a composite BB1 pulse \cite{Brown}. Ideally, this shows no dynamics, since the system is initialized in an eigenstate of $\mathcal{H}_{XX}$, making any effects of decoherence clearly identifiable. However, with the addition of a small $B^y$ field this system also exhibits domain-wall confinement \cite{Tan2019}, which can result in non-trivial, slow dynamics between the confined quasiparticles. As a result, we expect that this sequence provides a reasonable estimation of the coherence we could expect when studying an interesting many-body system.

As shown in Fig.~\ref{fig:Fig4}, dynamical decoupling again greatly helps to preserve this state. This can be quantified with the generalized imbalance $\mathcal{I}(t)$, a spin-spin correlator reflecting the memory of the initial spin configuration \cite{Morong2021a}:

\begin{align}
    \mathcal{I}(t)&=\frac{\sum_{j}\left\langle\sigma_{j}^{z}(t)\right\rangle\left(1+\left\langle\sigma_{j}^{z}(0)\right\rangle\right)}{\sum_{j}\left(1+\left\langle\sigma_{j}^{z}(0)\right\rangle\right)}\nonumber\\
    &-\frac{\sum_{j}\left\langle\sigma_{j}^{z}(t)\right\rangle\left(1-\left\langle\sigma_{j}^{z}(0)\right\rangle\right)}{\sum_{j}\left(1-\left\langle\sigma_{j}^{z}(0)\right\rangle\right)}.
\end{align}
$\mathcal{I}$ is a constant under the target Hamiltonian, while dephasing or thermalizing dynamics will both drive it towards zero.

Fitting $\mathcal{I}(t)$ to an exponential decay, the scaled coherence time without dynamical decoupling is $\overline{J_0}\tau=4.2$, while with dynamical decoupling it is extended to $\overline{J_0}\tau=22.1$, showing that the benefit of dynamical decoupling demonstrated in two-ion tests persists in a much larger system. Addition of the decoupling pulses significantly extends the timescales accessible in this simulation, potentially improving the ability to resolve slow dynamics.

\subsection{Decoupled Floquet Hamiltonian Engineering: Realizing the Haldane-Shastry model}
\label{sec:Haldane-Shastry}

\begin{figure*}[!htb]
\centering
\includegraphics[width= 0.75 \textwidth]{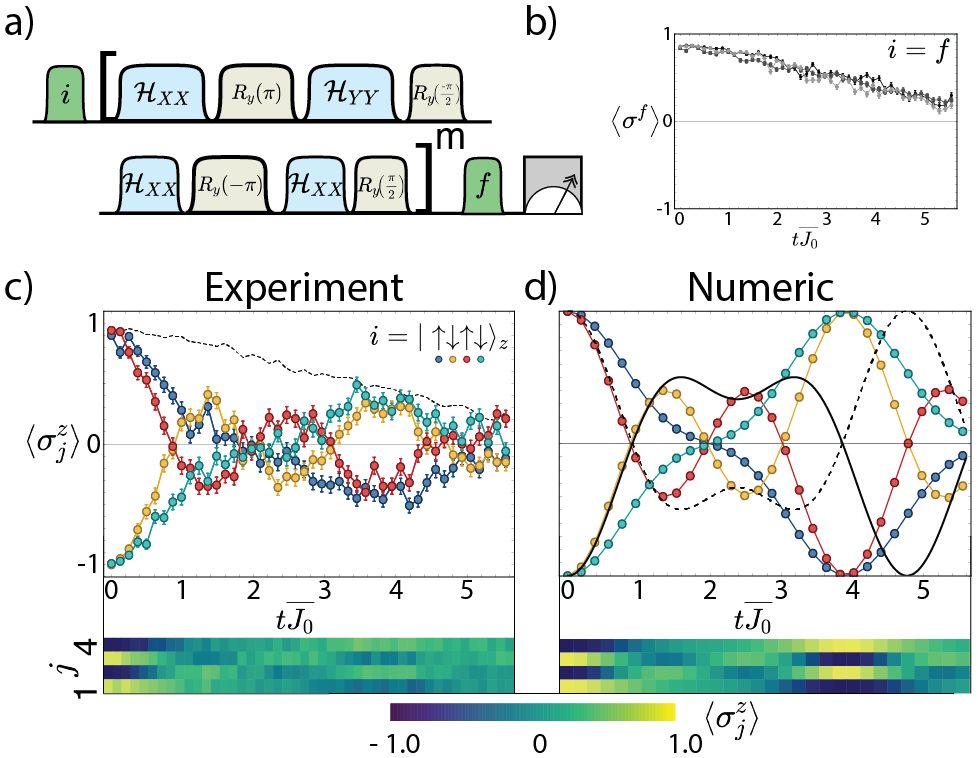}
\caption{Quantum simulation of the approximate Haldane-Shastry model using Floquet engineering. a) We prepare an initial state $i$, apply the decoupled Floquet engineering sequence (Eq.~\ref{eq:Heisenberg}) that results in $\overline{\mathcal{H}} \sim \mathcal{H}_{H-S}$, and measure in some basis $f$. b) Creating a polarized state and measuring it in the same basis, a comparable slow decay of the average magnetization is seen for $i$ along $x$, $y$, or $z$ (dark to light). c) Preparing a N\`{e}el state, the dynamics measured along $z$ (left) show complex oscillations and recurrences without thermalization. The average magnetization from b) (dashed line) is provided as a guide to the eye indicating the approximate rate of depolarizing decoherence. d) Numerics show very similar behavior as experiment up to the overall decoherence. Colored lines with markers (and lower heatplot) are a numerical simulation of the experimental Hamiltonian, while black lines (dashed for odd ions, solid for even ions) are the ideal, translationally symmetric Haldane-Shastry chain with the same number of ions and same average nearest-neighbor spin coupling. Experimental points are the average of 200 repetitions, with error bars of 1 s.e.m.}
\label{fig:Fig5}
\end{figure*}

As a final demonstration of these tools, we apply Floquet Hamiltonian engineering to a long-range Heisenberg model in the Haldane-Shastry regime. The Haldane-Shastry model \cite{Haldane1988,Shastry1988} can be formulated as a periodic spin-1/2 chain with long-range antiferromagnetic spin-spin interactions:

\begin{align}
    \mathcal{H}_{H-S}=\sum_{j<j'}\frac{J}{3|j-j'|^2}\vec{\sigma}_j\cdot \vec{\sigma}_{j'},
    \label{eq:HS}
\end{align}
with sites that are evenly spaced around a circle. This model has attracted extensive interest because it is exactly solvable using the asymptotic Bethe ansatz, and features a spin-liquid ground state and fractionalized spinon quasiparticles \cite{Greiter2019}. Previous work proposed to experimentally create this model with trapped ions \cite{Gra2014,Bermudez2017,Birnkammer2020} or atoms in a photonic crystal waveguide \cite{Hung2016}; however to our knowledge it has not been demonstrated.

We approximately engineer the Haldane-Shastry model using the dynamical decoupling sequence for a Heisenberg model (Eq.~\ref{eq:Heisenberg} and Fig.~\ref{fig:Fig5}a) with four ions whose spin-spin interactions obey an inverse square power law: $J_{j,j'}\approx J_0/|j-j'|^2$. Because the sequence now relies on a fast drive limit, we set $t_1 J_0=0.05$. This model is a convenient target for our technique because reaching inverse-square power law couplings requires a large detuning of $\delta/\eta\Omega=9.9$. At this large detuning, the scaled coherence time without dynamical decoupling is $\overline{J_0}\tau<1$, as suggested by Fig.~\ref{fig:Fig3}b, making decoupling necessary to see any nontrivial dynamics.

Our experimental realization deviates from the ideal Haldane-Shastry model in several ways. At the level of the Hamiltonian, the experiment has open boundary conditions, unlike the original version of this model, although open generalizations have also been previously studied \cite{Simons1994,Bernard1995}. Furthermore, the spin-spin couplings in the ion chain do not precisely obey an inverse square power-law dependence \cite{Monroe2019a}. For the data shown, the best-fit power law constant is 2.05, and the largest deviation of a coupling from the fit is 0.08 $J_0$. Beyond the Hamiltonian form, the experiment also has decoherence, which in this regime is primarily due to motional heating (see Appendix \ref{sec:AppendixNoise}). Averaging over the pulse sequence, this may be approximated as depolarizing noise (see Appendix~\ref{sec:AppendixHSDecoherence} for numerics incorporating decoherence). Despite these differences, we can experimentally and numerically observe key signs of proximity to the ideal Haldane-Shastry limit. Due to the Heisenberg symmetry,  any finite initial magnetization along each direction is conserved up to a slow decay set by the decoherence (Fig.~\ref{fig:Fig5}b). We estimate the coherence time by a fit to the average of these decays, giving $\overline{J_0}\tau\approx4.5$. Meanwhile, preparing another initial state, such as a N\`eel initial state along $z$, results in persisting and oscillatory non-thermalizing dynamics consistent with near-integrability (Fig.~\ref{fig:Fig5}c). Similar oscillations and recurrences are seen in numeric simulations of the experiment, which are performed by solving the Schr\"{o}dinger equation for $\overline{\mathcal{H}}$ using the experimental spin-spin couplings (Eq.~\ref{eq:FullMSHamiltonian}), and in simulations of the exact Haldane-Shastry dynamics (Fig.~\ref{fig:Fig5}d). We note that the inversion of the first recurrence, seen in experiment and experimental numerics but not the ideal Hamiltonian, is a boundary effect that is determined in an open chain by the parity of the number of ions. In additional data (see Appendix~\ref{sec:AppendixAdditionalHS} and \ref{sec:AppendixLevelStats}), we show that this behavior persists for an alternate initial state, and numerically study the level statistics of the approximate and exact Hamiltonians. Furthermore, utilizing the flexibility of Floquet Hamiltonian Engineering, we modify the driving sequence to engineer a less-symmetric thermalizing model, for which each of these signatures is lost: the experimental dynamics no longer conserve $\langle \sigma^z \rangle$ and no longer exhibit revivals (see Fig.~\ref{fig:FigHSSM}).

Approximate realization of the Haldane-Shastry model opens up a number of exciting possibilities. These include studies of prethermalization and relaxation in a near-integrable system \cite{Castro-Alvaredo2016, Mallayya2019, Lopez-Piqueres2021}, and of the thermodynamic consequences of the expected fractionalized quasiparticles \cite{Myers2021a}. We leave a full theoretical and experimental study of the trapped-ion implementation of this model to future work.

\section{Outlook}
\label{sec:outlook}

We have demonstrated a general strategy to extend dynamical decoupling to quantum simulators undergoing unitary many-body evolution. Applied to a trapped-ion experimental platform, we have shown that this technique may extend its coherence time and accessible parameter regime, and can be integrated with Floquet Hamiltonian engineering to further expand simulation possibilities. We expect that these decoupling strategies will be useful in systems with other noise sources as well. For example, our experimental qubit is insensitive to external magnetic fields, making decoherence from physical magnetic field fluctuations (as opposed to simulated magnetic fields from qubit energy shifts) negligible. However, these are a leading source of decoherence in other types of trapped ion experiments, and are amenable to the techniques developed here. These methods are also potentially relevant to other platforms with similar control, including quantum simulators based on neutral atoms \cite{Jepsen2020, Bluvstein2021, Periwal2021} or superconducting circuits \cite{Xu2018b}. Extensions of this approach, in general including rotations on individual spins rather than global rotations \cite{Frydrych2014,Hayes2014}, can be applied to decouple essentially any noise that can be modeled as arising from a slowly-varying Hamiltonian with undesired couplings, while preserving a target Hamiltonian of arbitrary symmetry properties.

Our results also help to clarify the limiting noise in various regimes of our simulator's operation, suggesting possible paths forward. With dynamical decoupling, our noise model (see Appendix~\ref{sec:AppendixNoise}) suggests that the limiting noise over most regimes of interest is slow drifts in the motional trap frequencies, possibly due to temperature fluctuations that shift the resonant frequency of the rf trap voltages or small shifts in the electronic grounding. Therefore, future improvements should target this noise, which we have not investigated as extensively as other error sources. We can further predict that operating at $\delta/\eta\Omega=3.5$, if we can reduce these errors arising from motional frequency drifts by about a factor of 5 so that they are below other error sources, we will be able to increase our coherence to $\overline{J_0}\tau\approx30$. In addition to hardware improvements, this will be facilitated by advances in automated and efficient re-calibration of experimental parameters, which are already in use in similar experiments \cite{Egan2021}. Once achieved, assuming that an experimental signal typically persists up to time $t=2\tau$, this would correspond to a time of $t\overline{J_0}\approx 60$ available for quantum simulation. 

While we have focused on demonstrations with small systems, the fundamental error sources we have studied couple to each ion equally, providing promising fundamentals for system size scaling. For sufficiently long ion chains, the decreasing axial frequency leads to an additional source of decoherence \cite{Cetina2022} that may ultimately require moving to a modular architecture \cite{Monroe2014}. However, even direct scaling has led to quantum simulations with up to about 50 ion chains \cite{Zhang2017,Joshi2022}. Maintaining this noise up to comparable sizes would enable our simulator to reliably investigate dynamics that are highly challenging to access either for classical simulation or near-term gate-based quantum computers \cite{Flannigan2022}.

\section{Acknowledgements}
We acknowledge helpful discussions with G. Pagano, A. Kyprianidis, and F. Liu. This work is supported by the DARPA Driven and Non-equilibrium Quantum Systems (DRINQS) Program (D18AC00033), the NSF Practical Fully-Connected Quantum Computer Program (PHY-1818914), the DOE Basic Energy Sciences: Materials and Chemical Sciences for Quantum Information Science program (DE-SC0019449), the DOE High Energy Physics: Quantum Information Science Enabled Discovery Program (DE-0001893), the DoE Quantum Systems Accelerator, the DOE ASCR Quantum Testbed Pathfinder program (DE-SC0019040), the DoE ASCR Accelerated Research in Quantum Computing program (DE-SC0020312), and the AFOSR MURI on Dissipation Engineering in Open Quantum Systems (FA9550-19-1-0399).

\section{Appendix}

\subsection{Experimental details}
\label{sec:AppendixExperiment}
Our apparatus has been described in previous recent works \cite{Kyprianidis2021,Morong2021a,Tan2019, Pagano2019}. For the data shown here, we create crystals of 2 to 11 ${}^{171}$Yb$^+$ ions in an rf Paul trap with anisotropic secular trapping frequencies of $\omega_{xy} \approx 2\pi\times$ 4.8 MHz and $\omega_z=2\pi\times$ 0.5 MHz. Pseudospin states are encoded in two ground-state hyperfine levels: $|F=0,m_F=0\rangle=|\!\! \downarrow\rangle_z$ and $|F=1,m_F=0\rangle=|\!\! \uparrow\rangle_z$. State initialization via optical pumping prepares all the spins in $|\!\! \downarrow\rangle_z$ with a fidelity greater than $0.99$ per ion. For the data shown in Fig.~\ref{fig:Fig1}, the average magnetization is read out using a photomultiplier tube (PMT), with a typical detection fidelity of 0.99 per ion. For all other data, individual spins are detected using fluorescent light imaged onto a charge-coupled device (CCD) sensor, with a typical detection fidelity of 0.97 per ion.

Coherent global operations are created using stimulated Raman transitions, driven between two optical frequency combs generated by a pulsed laser \cite{Hayes2010}, with typical $\pi$ times of $t_\pi=5$ $\mu s$. Interactions are generated with three frequency combs set up to drive two stimulated Raman transitions, creating the two tones for the M\o{}lmer-S\o{}rensen scheme. The resulting effective coupling parameters primarily depend on the laser power and detuning, while only weakly depending on system size. For the two-ion data presented, a typical spin-spin coupling rate is $J_0=2\pi\times$400 Hz ($\delta/\eta\Omega \approx 4$). For the multi-ion data presented in Fig.~\ref{fig:Fig4}, the typical energy scale is $J_0=2\pi\times$204 Hz ($\delta/\eta\Omega = 7.5$), and the spin-spin couplings approximately follow a power law dependence, $J_{j,j'}\approx J_0/|j-j'|^p,$ with $p=1.32$. For the four-ion Haldane Shastry data presented in Fig.~\ref{fig:Fig5}, $J_0=2\pi\times$84 Hz ($\delta/\eta\Omega = 9.9$) and $p=2.05$.

For the data in Sec.~\ref{sec:twoIonTests}, involving tests with two ions, we extract both $\tau$ and $\overline{J_0}$ self-consistently from the oscillatory dynamics, as described in the main text. For the multi-ion data in Sec.~\ref{sec:multiIonTests} and Appendix~\ref{sec:AppendixAdditionalHS}, there is no longer a single oscillation frequency at $2\overline{J_0}$. We therefore calculate $\overline{J_0}$  from calibrations of the experimental parameters and Eq.~\ref{eq:FullMSHamiltonian}.

\subsection{Dynamical decoupling with rotating-frame $B^z$ terms}
\label{sec:AppendixBz}
Our quantum simulator platform has the capability to implement an effective $B^z$ field through a rotating-frame transformation. To perform this transformation, the qubit transition is defined with an energy shift relative to the ``true'' energy splitting of the atomic states. Practically, this requires changing the frequencies of the M\o{}lmer-S\o{}rensen beams to $\pm\mu-2B^z$, making them symmetrically detuned from the redefined transition, while also calculating the phase for all coherent operations, normally defined in the rotating frame of the qubit transitions, with a corresponding shift of $-2B^zt$.

It is straightforward to combine this rotating frame with the toggling frame conditions for dynamical decoupling. As a concrete example, consider the CPMG sequence of Example 1, consisting of two periods of evolution with interactions separated by two $R_y(\pi)$ rotations. In the toggling frame of the rotations, the two periods of interactions alternate between $B^z$ and $-B^z$. Accordingly, $\phi(t)$ should alternate between $-2B^z t$ and $+2B^z t$, with boundary conditions that maintain phase continuity. Applying these conditions at each transition, we find that for the first part of each cycle, which begins at time $t_0+2(n-1)(t_1+t_\pi)$, the phase should be $\phi(t)=-2B^zt+2B^z(t_0+2(n-1)(t_1+t_\pi))$, while for the second half beginning at time $t_0+(2n-1)(t_1+t_\pi)$ it should be $\phi(t)=+2B^zt-2B^z(t_0+2n(t_1+t_\pi))$.

\subsection{Error analysis for the CPMG sequence}
\label{sec:AppendixError}
Here we provide a more detailed error analysis for the CPMG sequence shown in Section~\ref{sec:CPMG}, focusing on the degree to which dynamically decoupling suppresses the noise. We first separate the Hamiltonian terms into ideal (static) terms and the error:

\begin{align}
    \mathcal{H}_1&=\mathcal{H}_{01}+\mathcal{O}(t), \\
    \mathcal{H}_2&=\mathcal{H}_{02}+\mathcal{O}(t), \\
    \mathcal{H}_{01}&=\sum_{j<j'}J_{j,j'}\sigma_j^x \sigma_{j'}^x+\sum_j B^x\sigma^x_j+B^y\sigma^y_j+B^z\sigma^z_j\\
    &=\mathcal{H}_{t}, \nonumber\\
    \mathcal{H}_{02}&=\sum_{j<j'}J_{j,j'}\sigma_j^x \sigma_{j'}^x-\sum_j B^x\sigma^x_j+B^y\sigma^y_j-B^z\sigma^z_j, \\
    \mathcal{O}(t)&=\sum_j\epsilon(t)\sigma^z_j.
    \end{align}
The evolution operator for one cycle is:
\begin{align}
    U_1&=R_y(\pi)\mathcal{T}\left\{ e^{-i\int_{t_c}^{t_c+t_1}(\mathcal{H}_{02}+\mathcal{O}(t)) dt} \right \} \nonumber\\
    &\times R_y(\pi)\mathcal{T}\left\{ e^{-i\int_{t_a}^{t_a+t_1}(\mathcal{H}_{01}+\mathcal{O}(t)) dt} \right \} 
\end{align}
with $\mathcal{T}$ the time-ordering operator. The exact frame transformation allows this to be rewritten as:

\begin{align}
    U_1&=\mathcal{T}\left\{ e^{-i\int_{t_c}^{t_c+t_1}(\mathcal{H}_{t}-\mathcal{O}(t)) dt} \right \} \nonumber \\
    &\times \mathcal{T}\left\{ e^{-i\int_{t_a}^{t_a+t_1}(\mathcal{H}_{t}+\mathcal{O}(t)) dt} \right \}  \\
    =& \mathcal{T}\left\{ e^{-i\int_{t_c}^{t_c+t_1}(\mathcal{H}_{t}-\mathcal{O}(t)) dt} e^{-i\int_{t_a}^{t_a+t_1}(\mathcal{H}_{t}+\mathcal{O}(t)) dt} \right\}
\end{align}
At this point, we invoke the Magnus approximation, in which the average Hamiltonian is calculated in progressive orders of the commutator between terms, which we explicitly write up to second order:

\begin{align}
    &\mathcal{T}\left\{ e^{-i\int_{t_c}^{t_c+t_1}(\mathcal{H}_{t}-\mathcal{O}(t)) dt}\right \}=e^{-it_1\overline{\mathcal{H}}_c},\\
    &\overline{\mathcal{H}}_c=\overline{\mathcal{H}}_{c1}+\overline{\mathcal{H}}_{c2}+\cdots, \\
    &\overline{\mathcal{H}}_{c1}=\frac{1}{t_1}\int_{t_c}^{t_c+t_1}\mathcal({H}_{t}-\mathcal{O}(t))dt \\
    &\phantom{\overline{\mathcal{H}}_{c1}}=\mathcal{H}_{t}-\int_{t_c}^{t_c+t_1}\frac{\mathcal{O}(t)}{t_1}dt,
    \end{align}
    
    \begin{align}
    &\overline{\mathcal{H}}_{c2}= \nonumber \\
    &\phantom{\overline{\mathcal{H}}_c}\frac{1}{2it_1}\int_{t_c}^{t_c+t_1}\int_{t_c}^t \left[\mathcal{H}_{t}-\mathcal{O}(t),\mathcal{H}_{t}-\mathcal{O}(t')\right]dt'dt \\
    &=\frac{1}{2it_1}\int_{t_c}^{t_c+t_1}\int_{t_c}^t (\left[\mathcal{H}_{t},\mathcal{O}(t)\right]-\left[\mathcal{H}_{t},\mathcal{O}(t')\right])dt'dt
\end{align}

We can perform the same manipulations for the term beginning at $t_a$: $\mathcal{T}\left\{ e^{-i\int_{t_a}^{t_a+t_1}\mathcal{H}_{t}-\mathcal{O}(t) dt}\right \}=e^{-it_1\overline{\mathcal{H}}_a}$, and combine exponentials with the Baker-Campbell-Hausdorff (BCH) theorem:

\begin{align}
    U_1&\approx e^{-it_1\overline{\mathcal{H}}_{c}}e^{-it_1\overline{\mathcal{H}}_{a}}\\
    &= e^{-it_1((\overline{\mathcal{H}}_{c}+\overline{\mathcal{H}}_{a})-it_1[\overline{\mathcal{H}}_{c},\overline{\mathcal{H}}_{a}]/2+O(t_1)^2)}
\end{align}
Finally, the above expression reduces to:
\begin{widetext}
\begin{align}
    U_1&= \exp{\left(-it_1\left(2\mathcal{H}_{t} +(\overline{\epsilon}_c-\overline{\epsilon}_a)\sum_j\sigma^z_j  -\frac{it_1[\mathcal{H}_{t},\sum_j \sigma^z_j]}{2}(\overline{\epsilon}_a+\overline{\epsilon}_c+(\delta \overline{\epsilon}_c-\delta \overline{\epsilon}_a))+O(t_1)^2\right)\right)}
\end{align}
\end{widetext}
where we have defined variables related to the mean and fluctuations of $\mathcal{O}$ over a duration of $t_1$: $ t_1 \overline{\epsilon}_n=\int_{t_n}^{t_n+t_1}\epsilon(t)dt$ and $(t_1)^2\delta\overline{\epsilon}_n =\int_{t_n}^{t_n+t_1}\int_{t_n}^{t}(\epsilon(t)-\epsilon(t'))dt'dt$.

It can be seen from this expression that the lowest-order noise is caused by any drift over the experimental sequence, while higher-order contributions come from the commutator of the noise term with the target Hamiltonian. Both can be reduced by reducing $t_1$, while additional symmetrization of the pulse sequence can be used to partially cancel higher-order terms \cite{Hatano2005}.

\subsection{Drive error}
\label{sec:AppendixDriveError}
We benchmark our drive error by applying a given drive to a single ion initialized in $|\!\! \uparrow\rangle_z$ 500 times (for 1000 total $\pi$ pulses) and fitting the decay. For the XY sequence, the fit is consistent with no decay over this period, and the fit parameters naively imply a lower fidelity bound of $F>0.99997$ per drive period (two $\pi$ pulses). For the BB1 sequence, we measure the fidelity of a single driving period as $F=0.9998$. As a typical experimental time scan uses 30 drive periods, the drive error is therefore negligible when using either robust sequence.

\subsection{Noise model}
\label{sec:AppendixNoise}
\begin{figure}[!htb]
\centering
\includegraphics[width= 0.45 \textwidth]{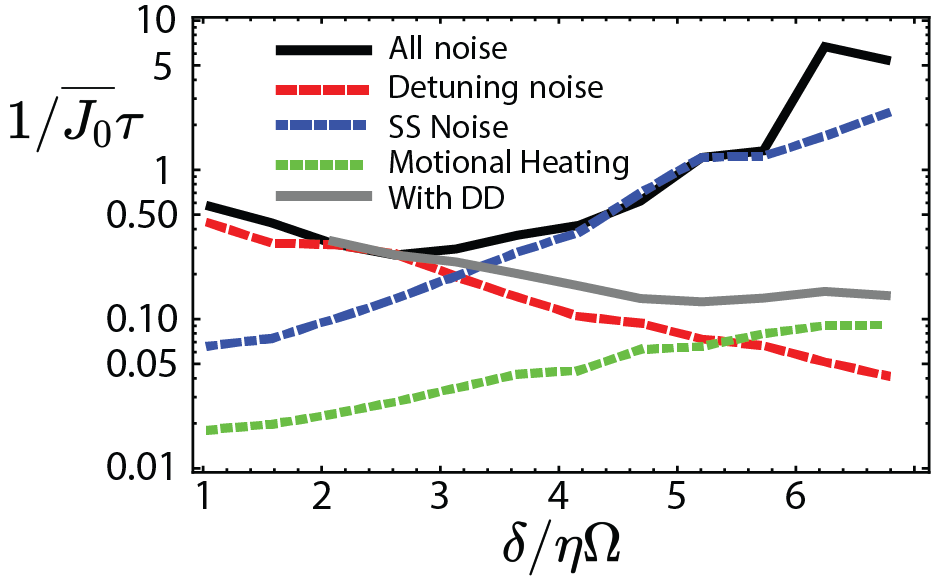}
\caption{Relative contribution of the error terms in our noise model. As $\delta$ increases, detuning noise decreases while Stark shift noise and motional heating errors increase. Dynamical decoupling effectively mitigates the Stark shift noise while having little effect on the other noise sources.}
\label{fig:FigNoiseModel}
\end{figure}

For the noise model used in Fig.~\ref{fig:Fig3}, we simulate the full M\o{}lmer-S\o{}rensen evolution (Eq.~\ref{eq:Molmer}) for both the red and blue tones, truncating the maximum number of phonons in each motional mode to two. The time step for the numerical evolution (using a Krylov subspace algorithm) is fixed at 15 ns. In addition, we add three types of experimental noise:

\begin{itemize}
    \item Stark shift noise. This consists of a random, independent intensity fluctuation on each of the two tones. These are constructed to have a power spectrum of $1/f^2$, over the bandwidth of [100 Hz, 1 MHz], and to have an overall fractional variation of $\sigma=0.021$ over the total sample time of 100 ms.
    \item Detuning noise. This consists of a Gaussian random variable, with $\sigma=4.5$ kHz, this is applied as a static shift to the detuning $\delta$ for each experimental run.
    \item Motional heating. This consists of a displacement operator with a random phase that is applied to the highest motional mode (the center-of-mass mode) every 1.5 $\mu$s, representing fast fluctuations of the global electric fields holding the ions \cite{Hempel2014}. The amplitude of these random displacements is set at 0.01 to match the decoherence observed in Fig.~\ref{fig:Fig3}. This results in a simulated heating rate of roughly 50 quanta/s. We can compare this to the rate observed in independent measurements, which is typically between 100 and 200 quanta/s, pointing to future opportunities to further refine our noise models with independent characterizations of the various processes.
\end{itemize}
We note that all of these are technical noise sources. The fundamental limit to coherent M\o{}lmer-S\o{}rensen evolution, off-resonant Raman scattering, is several orders of magnitude below these for the chosen parameters. The strengths of the Stark shift noise and detuning noise were taken to match the data in Fig.~\ref{fig:Fig3}, as were the exact power spectra over the experimental timeframe (choosing from possibilities of $1/f$ noise, $1/f^2$ noise, and an ensemble average of constant values). The strengths are similar to previous estimations of our error sources \cite{Pagano2019,Kyprianidis2021}, while the power spectra over experimentally relevant times are not strongly constrained by our previous measurements. Simulations were run for evolution times up to 0.5 ms, and averaged over 20 random realizations of all the error sources, before being fit to extract $\overline{J_0}\tau$. For numeric simulations of the dynamically decoupled sequence, the form of the experimental pulse shaping of the interactions was included, while we took the simplification of approximating the $\pi$ pulses as perfect and instantaneous.

Fig.~\ref{fig:FigNoiseModel} shows the relative contribution of each of these noise sources to $1/\overline{J_0}\tau$, as a function of the detuning $\delta/\eta\Omega$. Consistent with our expectations, the decoupled dynamics effectively suppress the Stark shift noise without similarly affecting the other sources.

\subsection{Haldane-Shastry numerics with decoherence}
\label{sec:AppendixHSDecoherence}

\begin{figure*}[!htb]
\centering
\includegraphics[width= 0.75 \textwidth]{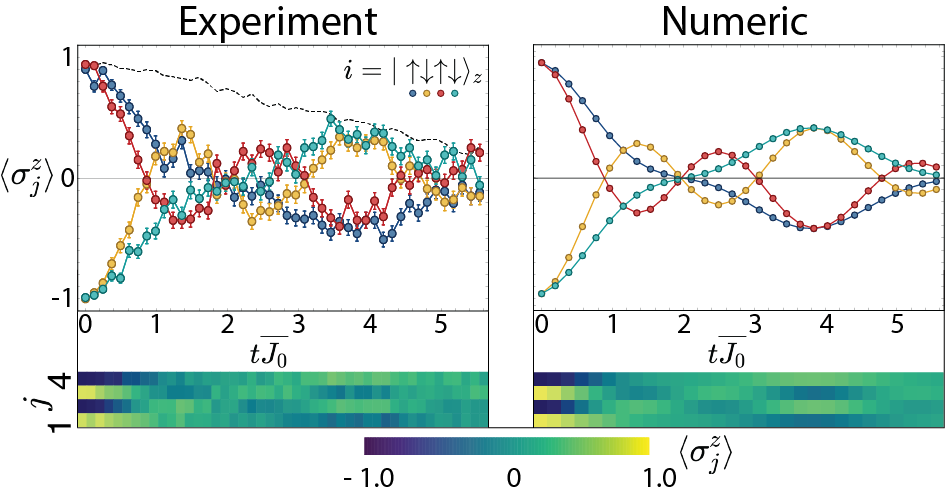}
\caption{Comparison of experimental data with a minimal decoherence model based on depolarizing noise with $\overline{J_0}\tau=4.5$. Inclusion of this effect results in numerics that have a strong agreement with experiment, suggesting that the primary influences on the experimental data have been accounted for.}
\label{fig:Fig5alt}
\end{figure*}

To support the comparison of experimental data with numerics, in Fig.~\ref{fig:Fig5alt} we present a comparison of the experimental approximate Haldane-Shastry data (reproduced from Fig.~\ref{fig:Fig5}) with numerics incorporating decoherence.

From the noise model described in the previous section, it can be seen that in the regime of the Haldane-Shastry data, $\delta/\eta\Omega=9.9$, for a decoupled simulation the dominant noise source is expected to be motional heating. This noise source can be treated as a stochastic $\sigma^x$ term \cite{Pagano2019}, but after considering the overall global rotations about the Bloch sphere in the Haldane-Shastry sequence the resulting decoherence is reasonably approximated as a simple depolarizing model. This leads the density matrix for the system to be approximated as $\rho = (1-p_1(t))I+p_1(t)|\psi(t)\rangle \langle \psi(t)|$, with $I$ the diagonal fully mixed density matrix, $|\psi(t)\rangle$ the desired evolution, and $p_1(t)=e^{-t/\tau}$.

We use the decay of the polarized data to estimate the rate of this decoherence process. Fitting the decay of the polarized data to a single exponential, we extract a coherence time of $\overline{J_0}\tau=4.5$. After applying the resulting correction to the numerics, the resulting dynamics are shown in Fig.~\ref{fig:Fig5alt}. Reassuringly, the numerics incorporating decoherence show close agreement with experiment.

\subsection{Modified Haldane-Shastry sequence}
\label{sec:AppendixHSmod}

\begin{figure*}[!htb]
\centering
\includegraphics[width= 0.75 \textwidth]{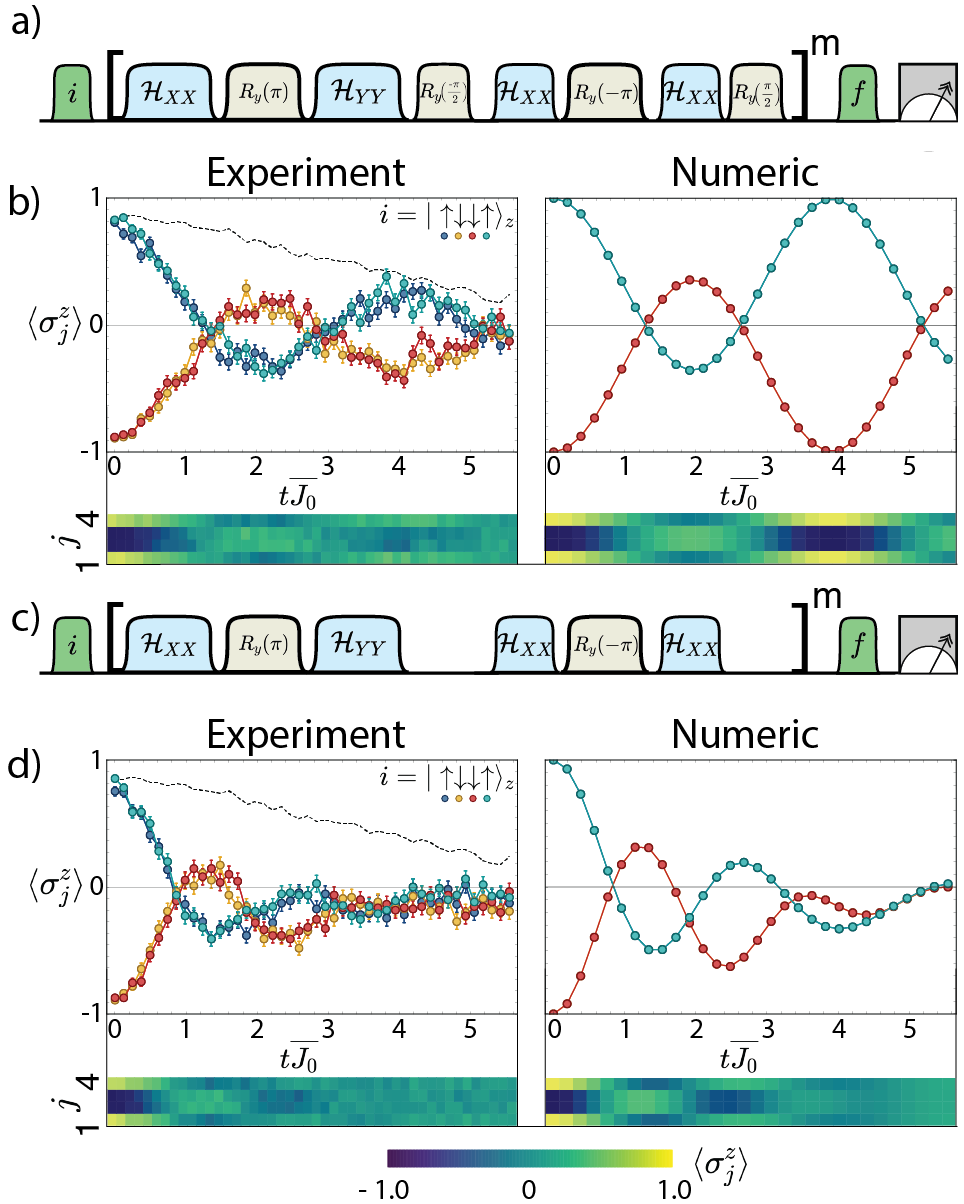}
\caption{Additional Haldane-Shastry data. a,b) Pulse sequence and data (left) compared to numerics (right) for the approximate Haldane-Shastry experiment, using a different initial state than in Fig.~\ref{fig:Fig5} that shows especially simple recurrences. c,d) Same as a,b but with a modified pulse sequence that breaks the symmetries of the Haldane-Shastry sequence (see text). The dynamics no longer exhibit the simple recurrences characteristic of near-integrability, and $\langle \sigma^z\rangle$ is no longer conserved. Experimental points are the average of 200 repetitions, with error bars of 1 s.e.m.}
\label{fig:FigHSSM}
\end{figure*}

\label{sec:AppendixAdditionalHS}
In Fig.~\ref{fig:FigHSSM} we show additional data for the Haldane-Shastry Hamiltonian engineering sequence. An additional initial state shows dynamics consistent with exact numerics, including a revival of the magnetization at late times.

To contrast with these results, and show the flexibility of our Hamiltonian engineering scheme, we also present data for a modified sequence in which two pulses are removed. This retains the decoupling of the Haldane-Shastry sequence, but results in an target Hamiltonian that has an anisotropic XY form:
\begin{align}
    \mathcal{H}_{HSmod}=\sum_{j<j'}\frac{J}{3|j-j'|^2}(2\sigma^x_j\sigma^x_{j'}+\sigma^y_j\sigma^y_{j'}).
    \label{eq:HSmod}
\end{align}
$\mathcal{H}_{HSmod}$ lacks the integrability of the Haldane-Shastry model, as well as the conservation laws for each spin projection. Correspondingly, we see dynamics that do not show simple oscillations with revivals or conserve total spin.

\subsection{Energy level statistics of Haldane-Shastry Hamiltonian}
\label{sec:AppendixLevelStats}

\begin{figure*}[!htb]
\centering
\includegraphics[width= 0.5 \textwidth]{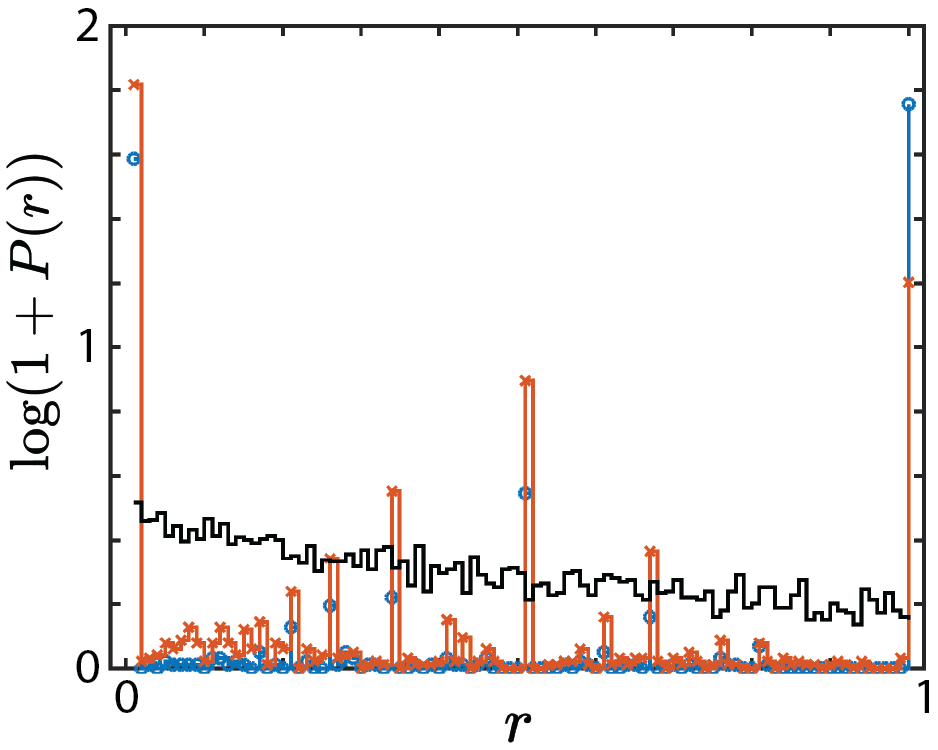}
\caption{Level statistics of experimental sequences and exact Haldane-Shastry Hamiltonian. Distribution of level spacing parameter $r$ (see text) for the exact Haldane-Shastry Hamiltonian (blue circles), the experimentally realized Hamiltonian (red x), and the modified sequence (Eq.~\ref{eq:HSmod}) (black lines). All results are computed using exact diagonalization for a twelve-ion system. While the modified sequence shows a relatively generic structure, both the experimental approximation to the Haldane-Shastry Hamiltonian and the exact Haldane-Shastry share a similar, highly non-generic and highly degenerate structure.}
\label{fig:FigED}
\end{figure*}

As described in the main text, our experimental Hamiltonian, while approximately realizing the Haldane-Shastry Hamiltonian, differs from it in several ways. While the exact Haldane-Shastry Hamiltonian can be described as spins arranged on a circle with couplings scaling as $1/r^2$, our experimental realization has open boundary conditions and couplings that are close to, but not precisely, a $1/r^2$ power-law dependence. In the main text, we have presented numerics for our specific experimental sequence to demonstrate that some of the key signatures of the Haldane-Shastry Hamiltonian, such as approximately integrable dynamics, are robust to these changes. However, this comparison is necessarily limited to the dynamics following certain initial states. To provide an alternate comparison reflecting the entire structure of the Hamiltonian, we show in Fig.~\ref{fig:FigED} the distribution of energy level spacings for the different Hamiltonians. Level spacings are quantified with the ratio of adjacent energy level gaps, defined as
\begin{equation}
r(n)=\frac{\text{min}(E_{n+1}-E_n,E_n-E_{n-1})}{\text{max}(E_{n+1}-E_n,E_n-E_{n-1})}.
\end{equation}
The statistics of $r$ are an established probe of the extent to which a Hamiltonian is chaotic, by capturing the structure of the eigenvalues. For example, a transition from $r$ obeying a Wigner-Dyson distribution to a Poissonian distribution is commonly seen for models undergoing a generic thermalizing to localized phase transition \cite{Oganesyan2007}. In the current application, the non-generic Hamiltonians can have levels that are exactly degenerate to within numerical precision, so we add an infinitesimal term to all energy level differences that is reduced until the results converge. Thus, indeterminate values appear as $r=1$.

In Fig.~\ref{fig:FigED} the distribution of $r$ is shown for three different cases:
\begin{itemize}
\item an exact realization of the Haldane-Shastry Hamiltonian (Eq.~\ref{eq:HS}),
\item the approximate realization of the Haldane-Shastry Hamiltonian in our experimental ion chain. This consists of the average Hamiltonian $\overline{\mathcal{H}}$ corresponding to the pulse sequence Eq.~\ref{eq:Heisenberg}, which describes a Heisenberg Hamiltonian with approximate $1/r^2$ couplings and open boundary conditions,
\item and $\overline{\mathcal{H}}$ for the modified pulse sequence which breaks the Heisenberg symmetry  (Eq.~\ref{eq:HSmod}).
\end{itemize}

For each, we numerically diagonalize a system of twelve ions, rather than the experimental four, so that there are enough eigenvalues to resolve the level structure. While this does have the effect of reducing the impact of boundary terms relative to the experimental data, it is nonetheless useful for understanding of the degree to which the class of Hamiltonians that we can experimentally create capture Haldane-Shastry physics.

At a qualitative level, it is clear that the Haldane-Shastry Hamiltonian shows a non-generic structure with peaks at certain spacings and a high degree of degeneracy ($r=0$ or 1). The experimental variant of this Hamiltonian shows a very similar structure, with shifted weights that reflect the boundary conditions and small integrability-breaking terms. Finally, the modified Hamiltonian is very different from the others, and has a more generic structure that is similar to other long-range spin Hamiltonians that are believed to be chaotic \cite{Morong2021a}. Thus, the level spacing structure is consistent with the dynamics in suggesting that key properties of the Haldane-Shastry model are resilient to the changes in our experimental approximation.

\bibliography{library.bib}

\clearpage

\end{document}